# Privacy Preserving Threat Hunting in Smart Home Environments

Ahmed M. Elmisery[1] and Mirela Sertovic[2]

[1] Faculty of Computing, Engineering and Science, University of South Wales, Pontypridd, UK

[2] Faculty of Humanities and Social Sciences, University of Zagreb
Zagreb, Croatia

ahmedmisery@gmail.com , msertovic@yahoo.com

**Abstract.** The recent proliferation of smart home environments offers new and transformative circumstances for various domains with a commitment to enhancing the quality of life and experience of their inhabitants. However, most of these environments combine different gadgets offered by multiple stakeholders in a dynamic and decentralized manner, which in turn presents new challenges from the perspective of digital investigation. In addition, a plentiful amount of data records got generated because of the day-to-day interactions between smart home's gadgets and homeowners, which poses difficulty in managing and analyzing such data. The analysts should endorse new digital investigation approaches and practices to tackle the current limitations in traditional digital investigations when used in these environments. The digital evidence in such environments can be found inside the records of log-files that store the historical events and various actions occurred inside the smart home. Threat hunting can leverage the collective nature of these gadgets, the vengeful artifacts observed on smart home environments can be shared between each other to gain deeper insights into the best way for responding to new threats, which in turn can be valuable in reducing the impact of breaches. Nevertheless, this approach depends mainly on the readiness of smart homeowners to share their own personal usage logs that have been extracted from their smart home environments. However, they might disincline to employ such service due to the sensitive nature of the information logged by their personal gateways. In this paper, we presented an approach to enable smart homeowners to share their usage logs in a privacy-preserving manner. A distributed threat hunting approach has been developed to elicit the various threat reputations with effective privacy guarantees. The proposed approach permits the composition of diverse threat classes without revealing the logged records to other involved parties. Furthermore, a scenario was proposed to depict a proactive threat Intelligence sharing for the detection of potential threats in smart home environments with some experimental results.

## 1    Introduction

The emergence of smart connected things that contain resource-constrained embedded systems with autonomous capabilities to enable connecting with other surrounding things and be self-aware of their internal states or external environment, has recently evolved into what is recently recognized as the internet of things. With the various wired and wireless communication technologies currently exist to facilitate connectivity between these heterogeneous devices, an increasing number of resource-constrained objects are getting connected with each other. These objects also have the capability of interacting with people, where they can collect data from people's everyday activities, then the collected data got exchanged with each other or remote services on the internet. Because of this feature, these objects are often titled "smart" and can be used as basic building blocks for smart automation systems. The smart home is one of the domains of IoT, that is composed of an IoT network of connected systems that facilitate connectivity between electronic sensors, analytical software and mechanical actuators inside the physical home environment [1]. The kind of setup, empower the smart homeowners with the ability to get notifications, to apply control and automate various activities performed within the home's perimeter. This also has an impact on enhancing the quality of daily chores from anywhere at any time through a smartphone application and an internet connection [2]. The progress of IoT-based smart home environments is currently gaining an increased momentum due to the wide range of improvements in the development of wireless protocols, embedded systems, cloud technologies, and the availability of internet-enabled smartphones. Most large businesses launching their own products to gain a share in this potential market and to inspire the advance of the next smart home ecosystem e.g., Nest smart thermostat, Apple HomeKit, Siri and Alexa enabled devices. These smart devices hold valuable digital data [3], potentially targeted by invasion attacks from external actors. Additionally, Smart home environments are frequently being targeted as well. It has been noticed a gradual increase in practices and available resources in relation to exploiting the vulnerabilities within smart home' devices. Since most of The IoT devices within Smart homes have a large number of vulnerabilities in their protocols, firmware, and software. Attackers are always ready to abuse these vulnerabilities to gather, alter or delete private information of smart-home owners or damage their IoT systems. The current security techniques are not 100% effective in the face of these increasing attacks. The threats dynamics for the smart home environment are very high, almost new vulnerabilities or unusual exploits are getting discovered daily in these systems. Data from different IoT systems of a smart home can be logged in different formats. From a threat hunting perspective, such logged data about various events in the IoT systems can give an indication about the behaviors and functionalities of these systems. In our work, we refer to IoT-based smart home environments as smart home.

Digital investigations are scientifically proven methods concerned with examination and analysis of digital evidence. The success of these tasks is highly depending on the ability of the forensic investigators to analyze large volumes of digital forensic data to locate suitable evidence. This, in turn, will require massive computational resources because of the size of the involved data. With the increasing cases of cybercrimes utilizing IoT devices [4], the need for applying digital investigations into IoT domain became indispensable. Since, the new IoT paradigm has been exposed to various vulnerabilities, which can induce a new type of cybercrimes that is accomplishable through these devices. However, it is necessary to adapt the processes of digital investigations when being applied into IoT scenarios, due to certain characteristics imposed by the IoT paradigm that make it different from other contexts, such as the increasing number of connected devices, the heterogeneity, diversity of collected data formats, and proprietary protocols of connected devices requires type-specific evidence retrieval tools, and the resource-constrained nature of these connected devices. Hence, existing digital investigations tools cannot successfully be applied to this paradigm without considering these features. IoT Forensics [5] is a relatively new term that characterizes a new paradigm in the digital investigations committed to implementing forensic practices for the collection and analysis of evidence from the peculiarities of Internet of Things (IoT) infrastructure. This new paradigm follows the well-established principles of traditional computer forensics, particularly targeting identification, acquisition, preservation, analysis, and reporting of digital evidence. Nonetheless, IoT forensics faces many challenges due to the inadequacy of currently usable computer forensics tools and methodologies in the IoT realms [6]. Furthermore, the tremendous amount of diversified data generated by the IoT devices poses a substantial challenge for the digital investigators to smoothly determine the precise portion of crucial data required for further analysis and examinations [7]. The potential evidence source in IoT forensics may include household appliances, health and fitness devices, entertainment systems, connected home hubs, Home monitoring/security systems, and outdoor gadgets, among others. While, in traditional computer forensics the potential evidence source can be computers, main servers, networking devices or mobile phones. In respect of the particular kind of the digital evidence that can be found, in IoT forensics, the digital evidence can exist in a standard or proprietary format, unlike the traditional computer forensics where the digital evidence mostly exists in standard formats. The last challenging factor in IoT forensics is the complex hardware architecture and diverse operating systems these gadgets have, along with monopolistic protocols and hardware that are subject to unclear standards.

One of the recent emerging themes in digital investigations is threat hunting, which is a preemptive cyber-defense activity that involves iterative searching through networks to detect sophisticated threats or potential vulnerabilities that elude existing security solutions, which differs from the current approaches that depend on passive waiting for cyber threats that might be detected if a violation in the previously configured conditions occurred inside any of the deployed network security solutions (IDS, IPS, SIEMS). The proactive approach can profoundly support the creation of a strong digital perimeter that continuously tracks pieces of evidence of new penetrative threats or malicious activities on the smart home environment. The threat hunting activity crucially relies on the cooperation between various groups of smart home's gadgets to repeatedly collect data to hunt for evidence regarding potential vulnerabilities without halting the functions of a smart home environment. The success of threat hunting activity highly depends on the ability of the system to execute threat-based security analysis approaches on large volumes of incomplete data from smart home's gadgets to locate suitable evidence, this, in turn, will require a lot of computational resources due to the volume of the data involved. The main objective of this research is to propose a collaborative threat hunting platform, which aims to share pieces of evidence with other smart home environments to facilitate prompt detection of attacks against the smart-home owners, or the infrastructure of the smart environment and for mitigation. The proposed platform can have a vast impact on alleviating the need for high computational resources, since only high confidence rules and artifacts related to attacks observed on other smart home environments will be shared. Furthermore, this approach can assist to overcome the complexity of evidence acquisition which might be incomplete or inaccurate when getting extracted from a single source. Finally, logs from several smart environments' gadgets collectively support each other since recording the same data in two different logs makes the extracted evidence stronger and can guarantee the genuineness of the threat to particular systems. An additional important requirement for threat hunting that the evidence data collected from various smart home environments are going to be processed collaboratively. Hence, scalability is essential in order to support the construction of large threat reputations groups with semantically enriched information. In order to prevent malicious entities from participating in the formation of threat reputations groups, tampering the results or compromising smart-home infrastructure to launch other attacks; every smart home environment needs to provide a set of credentials or prove identity during any of the steps required for the formation of threat reputations groups.

Most of the threat hunting solutions proposed so far have undervalued the demands to protect the privacy of end-users during the whole process of digital investigations. This situation has noticed real and significant implications for privacy concerns with the advent of IoT devices as an essential part of our daily activities as these gadgets are able to harvest and transmit personally identifiable information from every angle of our day-to-day activities. Thus, smart-home owners might be reluctant to participate in threat hunting activity, especially if they feel their own privacy is at stake by sharing data within their smart-home environment. These privacy issues that are related to threat hunting need to be taken into consideration for the advance of this kind of approaches. Most of privacy issues are related to how the data will be distributed and which laws govern the sharing of data. Thus, it is necessary to provide a technical assurance to smart-homeowners that privacy is guaranteed at all times and the sharing of information will not adversely affect their personal and professional reputation [8]. It is important that the proposed platform be able to enforce privacy principles in each of the phases of threat hunting approach and to provide initial functionality related to applying group-based access control and reputation mechanisms, to promote the establishment of trust between the various unnamed participants.

In this work, a threat hunting approach was proposed to implicitly elicit the relevant threat reputations groups from the multiple feeds of event logs. This process has been designed to be carried at the smart homeowner side. The presented system will attain security, privacy for event logs and will help smart homeowners to easily adopt a proactive approach for threat hunting. Any event logs collected from smart-home environments and shared with our system will be camouflaged using two-stage concealment protocols to preclude any potential risk of data breaches. This proposed approach also maintains the privacy of the data influx within the smart home such as sensitive usage patterns, events, and conditions, since any released data for threat hunting process will be concealed, and the original raw data will be stored in an encrypted form and only available for its owner. When the two-stage concealment pro-tcools handle the records within the event logs, any beneficial patterns for the threat hunting will be eliminated. Therefore, to support the analysis phase of threat hunting process on the concealed logs, some selective properties in the collected records have to be maintained to ease the ranking process related to threat reputations. This paper has been organized as follows, In Section 2, related works were summarized. Section 3 the proposed system that envisioned at the smart homeowner's side was outlined. The proposed adversary model was presented in section 4. Essential definitions related to the problem formulation along with the two-stage concealment protocols realized for the ranking process of threat reputations were explicated in section 5. Section 6, experimental results were depicted. Finally, the conclusions and future directions were given in Section 7.

## 2 Related Works

At present, digital investigation is a well-established technical aspect used by almost all big business. The processes of digital investigation begin only after a breach occurs. Lately, the attention has deviated from a reactive manner to a more proactive one, where security solutions will hunt for threats and vulnerabilities in the system to thwart breaches from occurring. The earlier detection of malicious activities or potential flaws, the better chance of reducing or avoiding future damages that may occur. This approach is what is known as threat hunting, which has recently emerged as a hot topic in the domain of cyber-security. However, there is a notable lack of literature review on this new approach. Each cyber-security vendor tends to promote its own definition of threat hunting to differentiate their own product as a threat hunting solution. In turn, this leads to ambiguity related to this concept. There are many different definitions to explain this term formulated based on its perception within the cyber-security community. For example, threat hunting can be defined as the process of seeking out adversaries before they can successfully execute an attack [9]. Sqrrl refers to threat hunting as a proactive and iterative searching through networks and datasets to detect threats that evade existing automated tools [10]. For the purpose of this research, threat hunting can briefly be defined as a preemptive activity that seeks for indicators of compromise in smart-home environments. The work in [11] proposed a framework that models multi-stage attacks in a way that describe the attack methods and the expected consensuses of these attacks. The groundwork of their research is to model behaviors using an Intrusion Kill-Chain attack model and defense patterns. The implementation of their proposed framework was employed using Apache Hadoop. The authors in [12] presented framework that utilizes text mining techniques to actively correlate information between the security-related events and the catalogue of attack patterns. The foundation of this work is to reduce analysis time and enhance the quality of attack identification. The work in [13] proposed a methodology that combines structural anomaly detection from information networks and psychological profiling of individuals. The structural anomaly detection uses graph analysis and machine learning to identify structural anomalies in various information networks, while the psychological profiling dynamically assembles individuals' psychological profiles from their behavioral patterns. The authors proceed to identify threats through by linking and ranking of the varied results obtained from structural anomaly detection and psychological profiling. In [14] an aspect of threat detection was introduced, which relies on identifying unexpected changes in edges' weights over time. Wavelet decomposition method was employed to differentiate the transient activity from the stationary activity in the edges. The authors in [15] proposed the usage of data mining techniques through visual graphical representation to overcome different threats. In their research, two new visualization schemes were proposed to visualize threats. In [16], a botnet detection framework has been developed, which utilizes cluster analysis to characterize similarity patterns of C&C communication and activities flows. The proposed framework sniffs the network traffic then executes two types of parallel analyses, one analysis is performed for detecting a cluster of hosts with similar communication patterns, and the other inspects the packet payloads to detect anomalous activities. The activities are later grouped to detect a cluster of hosts with similar malicious behavior. A cross-correlation process is utilized to merge the results of preceding analyses to produce meaningful groups of malicious hosts that might be forming a Botnet. The trade-off between digital investigation and privacy was discussed in [17], where the authors proposed privacy preserving forensic attribution layer to attain a balance between privacy and digital investigation processes. Group-based signatures were utilized as a part of their proposed solution to achieve the previous goal. Finally, the authors in [18] studied privacy issues in digital investigation and assumed that the forensics investigation process may violate the privacy of truthful users. The research work has proposed a protocol to offer privacy to these users while holding malicious users liable.

## 3      The Proposed System

The general approach of threat hunting is to log any and all events in its environment for further analysis. This allows forensic investigators to have more data during the correlation analysis of patterns in order to discover malicious behaviors. When the amount of collected data increases, this task becomes cumbersome and difficult to manage. Within smart home environments, the massive amount of generated data poses a restriction on any human-based analysis and calls for new automated approaches for threat hunting. In this work, clustering analysis will be utilized to infer potential future vulnerabilities and to rapidly label events on the smart home to cut down on detection time. The cooperative approach was utilized to get more threats related data and understand the whole context from nearby smart home environments. This in turn, vastly reduces the threat window which needs a further investigation. Moreover, the proposed platform actively mitigates these threats using immediate notification to the smart homeowner that contains the recommended remediation procedures along with automatically generating new detection and prevention rules for the deployed security solutions at smart homeowner's side. This offers a complete security solution that can be used to benefit new smart homeowners.

The research conducted on this paper aims to realize privacy by design approach [19] in which a middleware was proposed for governing the data privacy during the ranking process of threat reputations groups. The homeowners will not be forced to follow a binary subscription system, either to participate in the threat hunting by releasing their raw event logs or opt-out from the whole process. With the proposed system, the homeowners now have the capability to sanitize the sensitive information in any released event logs. In such a procedure, they will be empowered to disclose their data gradually. The proposed system can enable the homeowners to control their data that need to be shared with various threat hunting processes. Hence, they have the choice to enroll in any threat hunting group with a crafted version of their event logs. The key motivation behind our system is to apply a user-centric principle that follows the safest approach to preserve the sensitive records at the homeowners' gateways and to not release them in a raw form. Nonetheless, in order to participate in threat hunting and to obtain the related set of threat reputations groups, the homeowners should disclose their event logs in a specific way to facilitate the ranking process. Our proposed system relies on a middleware approach, that we named a cognitive middleware for cooperative threat hunting (CMCTH). Our middleware is composed of multiple cooperative agents and is being hosted in the smart home gateways. The collaboration between these different agents is essential to achieve privacy for the homeowners' data. Within CMCTH, every agent has a certain task; a local concealment agent executes a baseline concealment process which generates a sanitized event log by segregating the sensitive events based on end-user's policies. The local masking agent receives the sanitized event log then engages in the execution of the two-stage concealment protocols required for the ranking process of threat reputations groups.

The first stage protocol proposed in this research was entitled the secure threat ranking (STR), which is a distrusted cryptographic protocol, that is used to compose generalize virtual threat groups based on the individual event logs collected from different smart home systems. The second stage protocol was named the secure threat insights (STI), which takes as an input the virtual threat groups extracted from the first stage protocol then proceed to detect a set of real threat groups in every and all virtual threat groups. Aggregation topologies were employed in CMCTH to manage the data collection process from homeowners' gateways. The first stage protocol utilizes a simple ring formation topology for its processes and the second stage protocol uses a complex hierarchical formation topology for extracting precise real threat groups. The selection of these two aggregation topologies is adequate to the different steps in every protocol. Our proposed system relies on the existence of a centralized threat intelligence server (CTI). CTI is a centralized element on the proposed system responsible for initiating the ranking process and storing the final groups extracted after every run. CTI also offers a virtual workplace and a proxy for facilitating the interaction between smart homes' gateways of different vendors that might have uncommon communication protocols. The scenario presented in this research is the following: based on the several threat categories, CTI generate initial threat reputation groups then submit data related to these groups to the smart home's gateways to verify if the event logs might have occurrences of these threats. Every gateway is running its own CMCTH. CMCTH start the threat hunting process by inciting the local concealment agent to produce a sanitized event log that contains only generalized events that are related to their logs while segregating any sensitive events. Since this step cannot guarantee full privacy, CMCTH attempts to conceal these sanitized events by executing two-stage concealment protocols during the threat categorization process. The final threats ranking get offered based on the relevancy of the published events to different extracted threats. At the end of the threat hunting, CTI will permit the smart home's gateways to participate in a pertinent threat reputation group which will present a set of countermeasures to prevent these threats from occurring.

## 4　The Assumed Adversary Model

Our proposed system attains privacy for the sensitive records in the event logs. Every entity participating in the threat hunting process is following the semi-honest model. Hence, it is obliged to behave in accordance with the processes of the two-stage concealment protocols, but any intermediate data received from other entities could be stored for further investigation. In our model, we considered the centralized threat intelligence server (CTI) to act as an untrusted adversary that aims to gather the sensitive records in the event logs to be able to infer activities of various homeowners and to trace them back. We don't assume CTI to be an entirely malicious adversary which is a practical assumption. CTI is aiming to attain some business goals as well in order to boost their profits and reputations in the threat hunting industry. As a measure for the usefulness of our system, the achieved privacy is considered high only if the CTI can't infer activities of various homeowners from their event logs released for the threat hunting process.

## 5　The Proposed Concealment Procedure in CMCTH

We will start this section with outlining a set of significant notions used during this research work based on our previous research in [20, 21]. The event logs within our system are being presented in two forms, a sanitized event log and a sensitive event log. On one hand, the sanitized event log is a generalized version of the sensitive event log, where sensitive events get suppressed and other non-sensitive events get replaced with a set of hypernym phrases that are in the same semantic level of these non-sensitive events. The sanitized event logs can be considered the public data the smart homeowners agreed to disclose and released by CMCTH for the threat hunting process. On the other hand, the sensitive event logs stores records of personal events that smart homeowners avoid publishing in a raw form for other external entities. Privacy should be maintained when performing the ranking process related to threat reputations groups. Privacy should also be preserved when the final threats ranking get offered to a new entity on the system. The collected event logs used in threat hunting processes should also be preserved from CTI and/or any external entities involved in the whole procedure. In this paper the term "virtual threat group" can be defined:

**Definition 1.** A virtual threat group is the set $VC = \{RC_1, RC_2, \dots, RC_n\}$, where n is the number of real threat groups in $VC$, the virtual threat group has the following properties: (1) Each $\forall_{i=1}^n RC_i \in VC$ has a 3-element tuple $RC = \{I_{sg}, V_{sg}, d_{sg}\}$ such that $I_{sg} = \{i_1, i_2, \dots, i_l\}$ presents the set of sanitized events, $V_{sg} = \{v_1, v_2, \dots, v_k\}$ corre-sponds to the set of gateways, and $d_{sg} \in I_{sg}$ is the main- defining event of RC. (2) For every gateway $\forall_{i=1}^{l} v_i \in V_{sg}$, $v$ have the events $V_{sg}$. (3) $d_{sg}$ is the most frequent event in $V_{sg}$ events log, and this event considered as the "core-point" of this real threat group RG. (4) For any two real threat groups $RC_a$ and $RC_b$ ($1 \leq a, b \leq n$ and $a \neq b$) the following conditions are satisfied: $V_{sg_a} \cap V_{sg_b} = \emptyset$ and $I_{sg_a} \neq I_{sg_b}$.

### 5.1 The Two-Stage Concealment Protocols

This work presents two-stage concealment protocols that will be used for masking the event logs of smart homeowners when being released for threat hunting. The CMCTH is the element that is hosted on the gateways of smart homeowners and enforces the privacy preservation of the sensitive records on the event logs [22-29]. CMCTH will also execute the proposed cryptographic concealment protocols. The first stage concealment protocol was named secure threat ranking (STR) and the second stage concealment protocol was termed secure threat insights (STI). These protocols will facilitate the secure ranking extraction of threat groups from the masked event logs. These protocols will privately offer a ranked list of potential threats to new participants based on the relevancy of their released event logs to the different extracted threat reputation groups. Hence, any newly registered gateway can participate in any real threat group in a secure and private manner. The members of the same real threat group can share their systems configurations to prevent the occurrence of these threats and share information that can aid in handling a certain chain of threats facing their crucial systems.

### Secure Threat Ranking (STR) Protocol

The first stage concealment protocol aims to categorize different event logs into multiple virtual threat groups. CMCTH will face two challenges when categorizing those virtual threat groups. The first challenge is related to the representation of this threat group, i.e., an accurate intra-group closeness and a precise intra-group separation need to be clear in every extracted virtual threat group. The second challenge is related to attaining high privacy level for the sensitive records inside the event logs. Therefore, STR takes as an input the sanitized event log that was previously preprocessed by local concealment agent. This is a crucial step to maintain higher privacy levels for any published records. The sanitized event log is usually formed using public information by mapping any released event logs with a set of hypernym phrases obtained through taxonomy trees and/or public dictionaries that produce alternative events in the same semantic level of the original sensitive events. This process will form what is known as a sanitized event log as previously proposed in [20, 21] .
After generating sanitized event logs, CMCTH invokes the masking agent to run the distributed STR protocol to start building virtual threat groups based on the sanitized event logs that were submitted by smart homes' gateways. After running the STR protocol, every created virtual threat group will hold the set of smart homes' gateways who are largely sharing a similar set of events in their published event logs. The STR protocol gets executed in a distributed way. This protocol starts by organizing smart homes' gateways in a ring topology. STR protocol utilizes sanitized records published by gateway $V_c$ to create an event vector $V_c = (e_c(w_1), \ldots, e_c(w_m))$, such that m represents the number of unique events in the event log, and $e_c(w_1)$ describes the significance weight of such event $w_1$ in gateway $V_c$ (weighted frequency). The next computation steps utilizes the concept of term frequency inverse log frequency model that was presented in [30] as follows:

$$\text{Term} - \text{frequency}_{V_c}(w_i) = \#w_i \text{ in } V_c \log/\#\text{events in } V_c \log, \text{and}$$
$$\text{inverse} - \log - \text{frequency}_{V_c}(w_i)$$
$$= \log(\#\text{gateway}/\#\text{logs contain event } w_i), \text{where}$$
$$e_c(w_1) = \text{Term} - \text{frequency}_{V_c}(w_i) * \text{inverse} - \log - \text{frequency}_{V_c}(w_i)$$

The selection of similarity metric is an important step in STR protocol. Since an appropriate metric will be able to capture the hidden similarity between all sanitized records of every gateway. For this reason, we employed the Dice similarity metric. Let $V_c(V_d)$ to be two event vectors respectively for two gateways C and D then:

$$\text{GatewaysSimilarity}(V_c, V_d) = 2|V_c \cap V_d|/|V_c|^2 + |V_d|^2$$

In simple words, every two gateways C and D can be considered similar to each other if they are sharing many sanitized records with each other. Accordingly, the STR protocol should be able to infer that these two gateways should be belonging to the same virtual threat group. It is worth mentioning that any sensitive events will not be published and will be stored in an encrypted form at the smart homeowner side. The processes for the STR can be described as follows:

- For each threat hunting procedure, every two gateways $C, D \in V$ own a set of event vectors $e_c(w_i)$ and $e_d(w_i)$. Each one of them executes a predefined hash function denoted by h on its own set of event vectors to generate new hashed sets $V_c = h(e_c(w_i))$ and $V_d = h(e_d(w_i))$ respectively. *CMCTH* hosted on the gateway C will generate a two pair of encryption key E and decryption key U. *CMCTH* shares this encryption key E with the other gateway D. Computing the similarity between every two gateways is done by calculating two steps. The first one, is to compute the numerator, and the second one is to compute the denominator.

- For the correct execution of the STR protocol, one of the smart home's gateways should be selected as a trusted node for the aggregation process. Thereafter, topological ring formation is built between all of these gateways who decided to participate in the threat hunting in order to receive the calculated numerator values.

- The masking agent running on gateway D starts to hide $V_d$ by executing $B_d = \{e_d(w_i) \times r^D | w_i \in V_d\}$ where r is a random number for every event in its event log $w_i$. After finishing this step, the gateway D sends its $B_d$ to the gateway C.

- After the gateway C receives correctly $B_d$. The masking agent at gateway C starts to sign $B_d$ using its private key to obtain the signature $S_d$. Gateway C sends $S_d$ again to the gateway D in the same order as it has received it before.

The *CMCTH* running at the gateway D start the process of divulging the received event set $S_d$ by utilizing its r values to obtain the real signature $SI_d$ of the gateway D. the gateway C starts to implement the predefined hash function h on the real signature $SI_d$ to obtain the set $SIH_d = H(SI_d)$.

- The masking agent running on the gateway C also signs the event set $V_c$ to get the signature set $SI_c$. Additionally, masking agent implements the same predefined hash function h on the signature set $SI_c$ to form a new set $SIH_c = H(SI_c)$. The *CMCTH* on the gateway C submits the calculated values back to the gateway D.

- The masking agent at the gateway D initiates a comparison process between the two hashed sets $SIH_d$ and $SIH_c$ based on the knowledge it previously owns from $V_d$. The gateway D obtains the result of the intersection process between the two sets $IN_{C,D} = SIH_c \cap SIH_d$ which presents the interaction set between the two event sets of both of the gateways C, and D which will be denoted by $|V_c \cap V_D|$. *CMCTH* at the gateway D applies the predefined hash function h on the interaction set $IN_{C,D}$. After finishing the previous step, the gateway D encrypts with the public key of the trusted node the calculated hashed set along with the size of two sets $|V_D|$, $|V_C|$ and the gateways' pseudonyms identities. This encrypted data is later will be sent to the trusted node of this threat group.

- Finally, after collecting all of these intermediate results from every pairs of gateways at the trusted node. The trusted node starts the process of decrypting them, then after runs cluster analysis on these values using the S-seeds clustering algorithm [20] to order to obtain different virtual threat groups

The presented STR protocol performs all of these steps on $m$ hashed sanitized records that are distributed across $m$ parties without disclosing any of the raw values of these sanitized records.

**Secure Threat Insights (STI) Protocol**

The masking agent is the component within the *CMCTH* that is also responsible for implementing the second stage concealment protocol (STI protocol) on the extracted virtual threat groups obtained from the first stage concealment protocol (STR protocol). The idea of the STI protocol is to infer in a bilateral manner the set of correlated events exists between the sanitized records of event logs. The final output of this protocol aids in detecting the pertinent real-threat groups that exist within every virtual threat group. STI protocol is mainly based on our previous research work presented in [20, 21]. The main intuition of the STI protocol is to utilize the sets of frequent events that commonly exist between the event logs of multiple smart homes' gateways. If one of these frequent sets is large enough, a real threat group is formed that has this set a main/core topic. For the correct execution of the STI protocol, a topological hierarchical formation should be built between all of these gateways who decided to participate in the threat hunting. This topological formation aids in finding similar real threat groups spanned around the various virtual threat groups. The processes for the STI can be described as follows:

- The STI protocol is usually started after the termination of the first stage concealment protocol. However, it can also start by an indication from the CTI. The gateways in each virtual threat group confer together to elect one of them to act as a trusted node. The trusted node will be responsible for distributing its own catalog of 1-candidate frequent events. Upon receiving the 1-candidate frequent events, the gateways designate a local function hosted on the *CMCTH* to calculate their local frequent events on their sensitive event logs utilizing their own support and closure parameters. The algorithm discussed in [31] can run locally on every event logs to extract global & local frequent events of all event logs for the gateways within every virtual threat group.

- For the gateways in the same virtual threat group $\forall_1^n P_i$, the gateway $P_i$ start encrypting with its own key the locally extracted list of frequent events then sends this list to the second member $P_{i+1}$ in its virtual threat group, and so on for all gateways.

- This process got repeated for all gateways until the last gateway in the virtual threat group $P_{n-1}$ submits all the collected lists to the elected trusted node of this virtual threat group. The trusted node begins to calculate the global support for the global frequent events by simply aggregating all the received local supports from the different gateways. Moreover, the global closure for the global frequent events can also be calculated by finding the intersection between all the received local closures from the different gateways.

- The trusted node starts to encrypt and distribute the catalogs of global supports & closures in random order to the gateway $P_{n-1}$. The first gateway $P_{n-1}$ that receives these catalogs begins decrypting its own encrypted contribution using its own private key. After that, the gateway $P_{n-1}$ forwards these catalogs to another gateway $P_{n-2}$ also in random order. Finally, the trusted node gets back these catalogs, but this time these catalogs are only encrypted with the trusted node's own key. Therefore, final results can be generated.

- The trusted node initializes a real-threat group *RC* for every adjacent set of global frequent events. These initial real-threat groups comprise all gateways that have these global frequent events in their event logs. In the beginning, these initial real-threat groups could be overlapped between multiple virtual threat groups. However, the continuous progress of STI, they will be merged together such that every set of global frequent events will be representative for one real-threat group.

- For the event log of gateway $V_i$, the masking agent will need to assign an appropriate initial real-threat group $RC(c_i)$ by utilizing the following scoring function: $\text{SimilarityScore}(RC_i \leftarrow V_i) = \left[\sum_{w_i} e_r(w_i) * RC\_support(w_i)\right] - \left[\sum_{w_i'} e_r(w_i') * VC\_support(w_i')\right]$. Where $w_i$ represent the global frequent event in the event log r also this global frequent event is common in an initial real-threat group $RC_i$. The $w_i'$ is representing the global frequent event in the event log r and is not frequent in this initial real-threat group $RC_i$. After applying this scoring function on all event log of all gateways. At this point, every gateway will be able to determine its membership to an exactly one real-threat group. The representative of each real-threat group gets re-calculated based on the event logs of its current members.

- Inside every virtual threat group VC, the elected set of trusted nodes collaborate together to assemble a hierarchical structure of the real-threat groups, that have been extracted from the event logs of their members' gateway. Each real-threat group can now be represented using the set of global frequent k-events, in such case; it will act as a main/core topic or simply a representative. In the hierarchical structure, the real-threat group owns k-frequent events will be placed at level k of this structure. The parent of this real-threat group at level k-1 will own k-1 frequent events, which is also a subset of the frequent events owned by its child at level k. We have utilized the previously mentioned scoring function to derive the nominee parent for every child real-threat group. In the end of the step, the list of extracted real-threat groups got distributed between all the trusted nodes. This crucial step enables the integration of real-threat groups that own similar frequent events based on the inter-real threat group similarity and removes the restrained threat groups based on the intra- threat group separation. The proposed new similarity metric for this task is similar to the scoring function used in the STR protocol. The only new variation introduced in here in the normalization process that is employed to exclude the effect of the size of the real-threat group on the final result. This new utilized metric can be expressed as follows:

$$RC_{Similarity(RC_i \leftarrow RC_j)} = \left[ \frac{SimilarityScore(RC_i \leftarrow \forall_{x=1}^{n} V_x \in RC_j)}{\left[ \sum_{w_j} e(w_j) + \sum_{w'_j} e(w'_j) \right]} \right] + 1$$

The $Inter\ RC\_similarity\ (RC_i \leftrightarrow RC_j) = [RC\_Similarity(RC_i \leftarrow RC_j) * RC\_Similarity(RC_j \leftarrow RC_i)]$. Where $RC_i$ and $RC_j$ are two real-threat groups; $\forall_{x=1}^{n} V_x \in RC_j$ represents a single conceived event log for the real-threat group $RC_j$ that contain all the event logs of its current members. $w_j$ stands for a global frequent event exists in both of $RC_i$ and $RC_j$ while the $w'_j$ stands for a global frequent event exists only in the real-threat groups $RC_j$ but not in $RC_i$. Finally, $e(w_j)$ and $e(w'_j)$ represent the weighted frequency of both of $w_j$ and $w'_j$ in the real-threat group $RC_j$.

- Finally, when a new smart homeowner decides to participate in a threat hunting process, He/she invokes his/her own CMCTH to download from the CTI the catalog of core points for the available real-threat groups. Later, CMCTH begins the execution of the two-stage concealment protocols on the sensitive event logs stored at the smart homeowner's gateway. The similarity between the owner's sensitive event logs and the representatives of the real-threat groups can be computed locally at the homeowner's side. In the end, the CMCTH ranks all the similarity values then enroll its smart home's gateway in the real-threat group with the highest similarity value. the CMCTH will start receiving information related to the threat reputations of its owner usage events, the technical systems configurations of its owner will be automatically adjusted to prevent the occurrence of these threats and finally a set of threat analysis information will be provided to aid the smart homeowner in handling a certain chain of threats facing their crucial systems. Additionally, the smart homeowner can now get a detailed instruction related to mitigation and remediation procedures in the case of the occurrence of these threats. Finally, the CMCTH utilize the cooperative nature of the real-threat groups to automatically generate new detection and prevention rules for the current security solutions deployed at the smart homes.

## 6 Experiments and Results

The experiments presented in this research were performed on two Intel® machines connected using a local network. The Server has an Intel® Core i7 and the other machine has an Intel® Core 2 Duo. A data storage was done using a MySQL database for saving a set of event logs from various gateways. The CTI entity has been implemented as a web service. The *CMCTH* has been built as an applet to manage the various communications between the gateways themselves, and the interaction between the CTI and other gateways. The proposed two-stage concealment protocols were implemented using Java and boundycastel© library, RSA algorithm was used in encryption with key length of 512 for all experimental scenarios. The experiments have proceeded on a real smart home network that contains different IoT devices. In order to mimic a normal usage log, a dataset was pulled from an IPTV network which has been linked to another dataset containing 54 threat services of 30 IoT devices. The sanitized event logs were created based on recorded events. In order to measure the accomplished privacy level and the accuracy of results obtained using our system. We used precision and recall metrics as shown in Fig. (1). As noticed in the figure, good quality is attained when our solution identifies virtual threat groups first. These virtual threat groups will contain multiple sets of real-threat groups. With this way, our solution will be able to extract accurate information from the gateways who share the same data. Additionally, the impact of every sensitive event within the real threat group can be easily calculated. This will permit the *CMCTH* to extract and remove anomalies that are far from the common sanitized events.

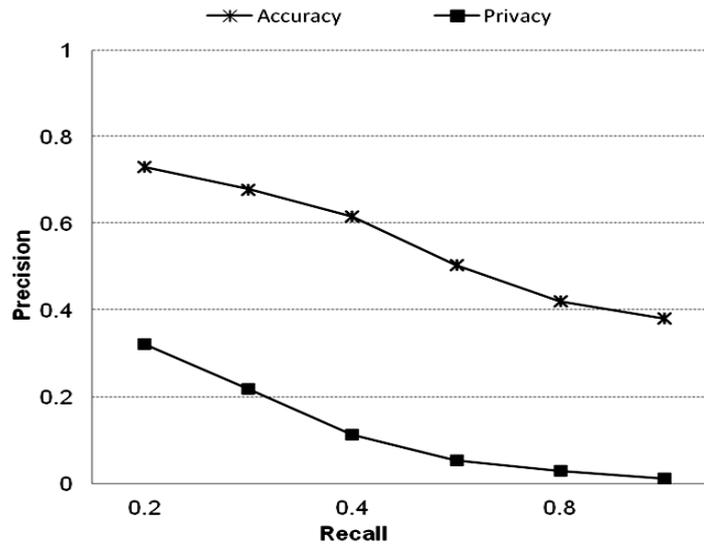

**Fig.1.** Accuracy and Privacy of the Extracted Threat Reputations Groups

We have also measured the impact of leaked sensitive events from different gateways on privacy levels during the run of the two-stage concealment protocols. We assumed a set gateways intentionally disclosed a portion of their sensitive events in the sanitized event logs released for threat hunting. For every set of sensitive events; the attack procedure mentioned in threat model was conducted with the aim of revealing other sensitive events stored in their sensitive event logs and based on the real-threat group they are belonging to. The obtained sensitive events were quantified and the results are shown in Fig. (1). As seen from the results, the proposed system can significantly reduce privacy leakages of the exposed sensitive events, However, the exposed sensitive events will only be hashed hypernym phrases that are in the same semantic level of non-sensitive events. The real sensitive events of other smart homeowners have already been suppressed from being released.

## 7    Conclusions and Future Directions

In this research work, a cognitive middleware for cooperative threat hunting (CMCTH) was presented which is hosted at the homeowners' gateways. CMCTH enroll the smart homeowners' gateways in a specialized threat hunting groups that permit the sharing of threat related information related to mitigation and remediation procedures in the case of the occurrence of these threats and also facilities the automatic re-configurations of smart homeowners' systems to prevent the occurrence of these threats. The formation of threat hunting groups is done without revealing any sensitive events logs to external entities. An overview of the two-stage concealment protocols was given. The performance of our solution was tested on a real dataset. The experimental results and the analysis clearly demonstrate that achieving higher privacy levels during the threat reputations processes is possible using our solution without the need to reduce the accuracy of the extracted threat reputations. A future research agenda for this research will include utilizing game theory to better compose virtual threat groups, multiple events publications and its impact on the privacy of the smart homeowners.

## References

1. Y. Seralathan, T. T. Oh, S. Jadhav, J. Myers, J. P. Jeong, Y. H. Kim, and J. N. Kim, "IoT security vulnerability: A case study of a Web camera." pp. 172-177.
2. A. Boztas, A. Riethoven, and M. Roeloffs, "Smart TV forensics: Digital traces on televisions," Digital Investigation, vol. 12, pp. S72-S80, 2015.
3. C. Gao, V. Chandrasekaran, K. Fawaz, and S. Banerjee, "Traversing the Quagmire that is Privacy in your Smart Home." pp. 22-28.
4. K. Biswas, and V. Muthukkumarasamy, "Securing smart cities using blockchain technology." pp. 1392-1393.
5. P. C. C. Shin, R. Liu, S. J. Nielson and T. R. Leschke, "Potential Forensic Analysis of IoT Data: An Overview of the State-of-the-Art and Future Possibilities," in 2017 IEEE International Conference on Internet of Things (iThings) and IEEE Green Computing and Communications (GreenCom) and IEEE Cyber, Physical and Social Computing (CPSCom) and IEEE Smart Data (SmartData), Exeter, 2017.

6. J. H. Ryu, Pradip Kumar Sharma, Jeong Hoon Jo, and Jong Hyuk Park, A blockchain-based decentralized efficient investigation framework for IoT digital forensics, pp. 1-16, 2019.

7. O. M. Adedayo, "Big data and digital forensics," in 2016 IEEE International Conference on Cybercrime and Computer Forensic (ICCCF), Vancouver, BC, Canada, 2016, pp. 1-7.

8. K. Australia, Cyber Threat Intelligence and the Lessons from Law Enforcement: KPMG Australia, 2015.

9. N. Lord, "What is Threat Hunting? The Emerging Focus in Threat Detection," Digital Guardian, 2018.

10. Sqrrl. "Cyber Threat Hunting," www.sqrrl.com.

11. P. Bhatt, E. T. Yano, and P. Gustavsson, "Towards a framework to detect multi-stage advanced persistent threats attacks." pp. 390-395.

12. N. Scarabeo, B. C. Fung, and R. H. Khokhar, "Mining known attack patterns from security-related events," PeerJ Computer Science, vol. 1, pp. e25, 2015.

13. A. G. Mahyari, and S. Aviyente, "A multi-scale energy detector for anomaly detection in dynamic networks." pp. 962-965.

14. B. A. Miller, M. S. Beard, and N. T. Bliss, "Eigenspace analysis for threat detection in social networks." pp. 1-7.

15. A. K. Bhardwaj, and M. Singh, "Data mining-based integrated network traffic visualization framework for threat detection," Neural Computing and Applications, vol. 26, no. 1, pp. 117-130, 2015.

16. G. Gu, R. Perdisci, J. Zhang, and W. Lee, "Botminer: Clustering analysis of network traffic for protocol-and structure-independent botnet detection," 2008.

17. M. Afanasyev, T. Kohno, J. Ma, N. Murphy, S. Savage, A. C. Snoeren, and G. M. Voelker, "Privacy-preserving network forensics," Commun. ACM, vol. 54, no. 5, pp. 78-87, 2011.

18. G. Antoniou, L. Sterling, S. Gritzalis, and P. Udaya, "Privacy and forensics investigation process: The ERPINA protocol," Computer Standards & Interfaces, vol. 30, no. 4, pp. 229-236, 2008.

19. I. S. Rubinstein, "Regulating privacy by design," Berkeley Technology Law Journal, vol. 26, no. 3, pp. 1409-1456, 2011.

20. A. M. Elmisery, K. Doolin, and D. Botvich, Privacy Aware Community based Recommender Service for Conferences Attendees: IOS press, 2012.

21. A. M. Elmisery, K. Doolin, I. Roussaki, and D. Botvich, "Enhanced Middleware for Collaborative Privacy in Community Based Recommendations Services," Computer Science and its Applications: CSA 2012, S.-S. Yeo, Y. Pan, S. Y. Lee and B. H. Chang, eds., pp. 313-328, Dordrecht: Springer Netherlands, 2012.

22. F. Beil, M. Ester, and X. Xu, "Frequent term-based text clustering," in Proceedings of the eighth ACM SIGKDD international conference on Knowledge discovery and data mining, Edmonton, Alberta, Canada, 2002, pp. 436-442.

23. M. Fung B. C, "Hierarchical document clustering using frequent item sets," Master's Thesis, Simon Fraser University, 2002, 2002.

24. A. M. Elmisery, S. Rho, and D. Botvich, "Privacy-enhanced middleware for location-based sub-community discovery in implicit social groups," The Journal of Supercomputing, vol. 72, no. 1, pp. 247-274, 2015.

25. A. M. Elmisery, S. Rho, and D. Botvich, "Collaborative privacy framework for minimizing privacy risks in an IPTV social recommender service," Multimedia Tools and Applications, pp. 1-31, 2014.

26. A. M. Elmisery, "Private personalized social recommendations in an IPTV system," New Review of Hypermedia and Multimedia, vol. 20, no. 2, pp. 145-167, 2014/04/03, 2014.

27. A. Elmisery, and D. Botvich, "Enhanced Middleware for Collaborative Privacy in IPTV Recommender Services " Journal of Convergence, vol. 2, no. 2, pp. 10, 2011.
28. A. M. Elmisery, and D. Botvich, "Agent Based Middleware for Maintaining User Privacy in IPTV Recommender Services," Security and Privacy in Mobile Information and Communication Systems: Third International ICST Conference, MobiSec 2011, Aalborg, Denmark, May 17-19, 2011, Revised Selected Papers, R. Prasad, K. Farkas, A. U. Schmidt, A. Lioy, G. Russello and F. L. Luccio, eds., pp. 64-75, Berlin, Heidelberg: Springer Berlin Heidelberg, 2012.
29. A. M. Elmisery, and D. Botvich, "An Agent Based Middleware for Privacy Aware Recommender Systems in IPTV Networks," Intelligent Decision Technologies: Proceedings of the 3rd International Conference on Intelligent Decision Technologies (IDT' 2011), J. Watada, G. Phillips-Wren, L. C. Jain and R. J. Howlett, eds., pp. 821-832, Berlin, Heidelberg: Springer Berlin Heidelberg, 2011.
30. F. Sebastiani, "Machine learning in automated text categorization," ACM Comput. Surv., vol. 34, no. 1, pp. 1-47, 2002.
31. D. W. Cheung, J. Han, V. T. Ng, A. W. Fu, and Y. Fu, "A fast distributed algorithm for mining association rules," in Proceedings of the fourth international conference on on Parallel and distributed information systems, Miami Beach, Florida, United States, 1996, pp. 31-43.